\documentstyle[12pt]{article}
\def\baselinestretch{1.8}
\begin{document}
\renewcommand{\baselinestretch}{1.5}
\title{Can a Logarithmically Running Coupling \\Mimic
a String Tension?}
\author{Michael Grady\\
Department of Physics\\ SUNY Fredonia, Fredonia, NY 14063}
\maketitle
\thispagestyle{empty}
\renewcommand{\baselinestretch}{1.8}
\begin{abstract}
\renewcommand{\baselinestretch}{1.8}
It is shown that a Coulomb potential using a running
coupling slightly modified from the perturbative form can
produce an interquark potential that appears nearly linear
over a large distance range. Recent high-statistics
SU(2) lattice gauge theory data fit well to this potential
without the need for a linear string-tension term. This calls into
question the accuracy of string tension measurements which are based
on the assumption of a constant coefficient for the Coulomb term.
It also opens up the possibility of obtaining an effectively confining
potential from gluon exchange alone.
\end{abstract}
\newpage
It is surprising that the interquark potential for pure-gauge SU(2)
and SU(3) lattice gauge theories fits as well as it does to a simple
linear+Coulomb law. Even the extremely high statistics SU(2) results
of the
UKQCD collaboration at $\beta=4/g^2=2.85$, which includes distances to
$R/a=24$ and with a relatively small physical lattice spacing, $a$,
($a^{-1}\simeq 6.56$ GeV) which probes well both short and
long distances, requires no additional terms to fit the
data \cite{UKQCD}.
What is surprising about this is that one of the most
definite predictions
of perturbation theory, backed up by high-energy
scattering experiments,
is that
the effective coupling is a running coupling,
so one would expect the coefficient of the Coulomb term,
which can be taken to be a renormalized coupling, to depend upon
distance. At weak couplings corresponding to
short distances this should match
the logarithmic dependence given by renormalization group improved
low-order perturbation theory. If lattice gauge theory is to
be successfully
matched onto perturbation theory, then at least the
short distance part of the potential should be allowed to run.
This has been tried
and gives indications of a reasonable match to perturbation
theory \cite{Michaeletc}.
For longer distances (say $R/a \simeq 6$ on the above lattice) the
coupling is generally assumed to stop running, to allow an accurate
determination of the string tension. However, the fits which show
a running coupling at shorter distances show no indication that
the running
is slowing down. The stopping of the running
coupling has been justified by the strong-coupling string model
of L\"uscher
which predicts the coefficient of $1/R$ in the potential to be the
constant value of $\pi/12$ \cite{Luscherold}.
The problem
with this is that the couplings for which this
string picture become valid are
probably much stronger than those of the
simulations being discussed here \cite{Gutbrod}.
There is very little independent evidence for the stopping
of the running
of the renormalized coupling. In fact, this would seem
to contradict scaling
studies based on Wilson loop ratios
which show that the bare coupling continues to
run for all values of the coupling.
One cannot have a fixed point in the $\beta$-function
for the renormalized coupling,
which the stopping of the running would imply, without
a corresponding
fixed point in the bare coupling $\beta$-function.

In this letter I investigate the effects of relaxing the assumption
that the coupling stops running at long distances.
It is shown that even the one-loop
perturbative potential roughly mimics the linear+Coulomb form
over the large distance range included
in the aforementioned simulation.
That is, the rising running coupling
can temporarily counteract the falling Coulomb law to
produce an approximately
linear potential over a certain range. In the end the one-loop
perturbative potential is ruined by the Landau pole singularity.
A phenomenological potential consistent with
perturbative functional forms is introduced which contains
an additional
piece which can kill the Landau pole. This form is
found to fit the
data every bit as well as the Coulomb+linear form. The potential
eventually falls off the linear trend, but not until $R/a\simeq 50$,
well beyond the range of the simulation or planned simulations for
some time to come.
Thus it will be shown that a logarithmically running coupling is
capable of producing a phenomenologically confining potential. The
potential is not absolutely confining, however it should
be remembered
that this is not necessary in
the real world with light quarks, since beyond a certain separation
a meson pair will form, breaking the ``string''.
Thus the pure gauge simulation is only relevant to real world
physics up to
a distance of order 1~fm.

In what follows, the focus is first shifted to the
interquark force, as
opposed to the potential, because it is more easily compared to the
perturbative result \cite{Creutz,Billiore}.
For lattice SU(2) the magnitude of the force may be written
\begin{equation}
F(R)=(1+3a^2 /4R^2)3\alpha(R)/4R^2
\end{equation}
The prefactor in parentheses is from an approximation to the lattice
Coulomb propagator (valid for $(R/a)\ge 2$) which differs
slightly from that
of the
continuum for small $R/a$ \cite{Gutbrod}. It actually has very
little effect on the fits.
The potential is taken to be the integral of the force, which may
become complicated for complicated $\alpha(R)$.
For this reason it is
much easier to work with the force.
The main disadvantage to working with the force is that the
Monte Carlo data give more directly the potential itself.
Data for the force can be obtained from the potential
data by taking finite differences. This
introduces some horizontal(i.e. $\Delta R$) uncertainty into
the data, since one
is not sure where within the $R$-interval to plot the
force value against.
Usually the midpoint is chosen, but because of the
inverse relationship here
it is better to use the geometric mean, which gives exact results for
a $1/R$ potential. For example, the force derived from the
difference between
$V(4a)$ and $V(2a)$ is plotted
against $R/a=\sqrt{2\cdot 4}=2.83$.

Fig.~1 shows a fit to the
UKQCD data from Ref.~[1] for $R/a\ge 4$ (the $R/a=2$
point was also excluded from the fits in Ref.~[1]). The
short-dashed line is a one-parameter fit to
the one-loop renormalization-group improved force with
\begin{equation}
\alpha (R)=\left( 1+8\pi b_0 \ln(R_0 /R) \right) ^{-1}
\end{equation}
where $b_0 =11/24\pi^2$. The fit gives $R_0 /a$=12.28.
Considering that it includes only effects from
(summed)
lowest order perturbation theory, it
falls suprisingly close to the data even in the high-$R$
region. If $b_0$ is
allowed to be a free parameter, only a marginally
better fit is produced
with $b_0$ increasing 5\% over its perturbative value. Such a fit
can be thought
of as a determination of the scaling parameter $b_0$ directly
from the data.
Thus the hypothesis that it is the running
coupling which is responsible for the form of the
force even at intermediate
and large distances appears reasonably consistent
with asymptotic scaling.
Note that the force function produced by the running coupling
is relatively flat in the region $13 \leq R/a \leq 22$,
varying only 10\%
from a constant. Thus the force in this region is much closer
to a constant force, which would mimic a
confining string tension, than
it is to a $1/R^2$ Coulomb force. The Coulomb
decrease is nearly matched by
the increase in effective coupling.

Although this function roughly fits the data it does not fit
this high statistics data well enough in detail, giving a
$\chi ^{2}/$DF$=4.4$. Clearly the fit is spoiled by
the appearance of the
Landau pole at $R/a\simeq 28$. This is widely believed to
be an unphysical result due to the partial summation of the
series.
   The fit to the two-loop renormalization group
improved force, using
the running coupling
\[
\alpha (R) = \left( 4\pi b_0 \left[\ln ((R_0/R)^2)
+(b_1/b_{0}^{2})\ln \ln((R_0/R)^2)\right] \right) ^{-1}
\]
with $b_1/b_0^2=102/121$ gives the large dashed line, which is
clearly worse than
the one-loop fit. This one-parameter fit gives $R_0 /a = 32.5$.
If one allows
the constants to deviate from their perturbative values,
not much improvement
results. The worse fit to the two-loop force can be
attributed to the even
earlier appearance of the Landau pole, around $R/a =24$. There
is still a relatively
constant region from about $10 \leq R/a \leq 20$, but the
value is about 50\%
too low. One can understand why the two-loop result might be worse
at long distances by considering the renormalization-group
$\beta$-function. If $\alpha(R)$ is to remain finite, the
true $\beta$-function,
which diverges from the axis like $g^3$ at small $g$, must
eventually turn
back toward the axis. The two-loop $\beta$-function
adds a $g^5$
term of the same
sign as the $g^3$ term, which causes the $\beta$-function
to diverge even faster.
This will cause it to disagree even worse with the
assumed behavior
of the true $\beta$-function at
large $g$, even though it is more accurate
than the one-loop result at
small $g$.

If the Landau pole could somehow be removed, then it would
not take a running coupling much different from the one-loop form
to fit the data, because the one-loop force already gives a
qualitatively successful fit. The approach to be taken is that
of a phenomenological extension of the one-loop force, consistent
with functional forms of (summed) perturbation theory.
It is reasonable to assume that
$\alpha (R)$ is given by a power series in $\ln (R_0 /R)$, but with
coefficients that differ from the simple geometric
series of the one-loop
bubble diagrams. A reasonable generalization for the
geometric series
summation is that of the Pad\'e approximate:
\[
\alpha (R)= \frac{\sum_{j=0}^{n} a_j \left( \ln (R_0 /R) \right) ^j}
{\sum_{k=0}^{m} c_k \left( \ln( R_0 /R) \right) ^k}
\]
Excellent agreement with the data is found with a one-term
extension to
the one-loop form, namely $n=0$, $m=2$. The solid
line in Fig.~1 is a fit
to the force associated with the
phenomenological running coulpling:
\[
\alpha (R) = \left( 1+8 \pi b_{0} c \ln (R_{0} /R) +
d (\ln (R_{0} /R))^2 \right) ^{-1}
\]
where the three parameter fit gives $c=1.1163$, $d=0.4856$,
and $R_0=10.953$.
The fit is extrapolated beyond the data to show a large
region of approximately
constant force extending at least to $R/a \simeq 35$. This is,
for all practical purposes, indistinguishable from the
force due to a string tension. The force does
eventually fall off around
$R/a =50$,  but this region is well beyond
the reach of today's simulations. It is remarkable
that a two-term
logarithmic form could so effectively kill the $1/R^2$
Coulomb dependence
over such a large distance range.  The fit
has a $\chi^2/$DF $= 0.7$, compared
with 0.9 reported for the Coulomb + linear fit in Ref.~[1].
Note that the
point at $R/a = 2.83$ is not included in the fits. Vertically it
is quite far off the
fit, but a small horizontal correction would easily place it on.
Note also the close
agreement with the two-loop result at small $R$.
This was not in any way built
in, but came out as a result of the fit. The agreement with the
two-loop result is quite close in the
range $2.5\leq R/a \leq 4$. The
phenomenological force then begins to
disagree with the two-loop force
again for smaller $R$. Since there is no data
in this region, it is not
an issue for the present fit, but for work which
includes shorter distances,
it would probably be best to match the new $\alpha (R)$
onto the two-loop
result in this region of approximate agreement
and to use the two-loop
$\alpha (R)$ for smaller distances than this, because it is almost
certainly valid here.  This would also cure the new
$\alpha (R)$ of its major flaw, namely that it does not obey the
perturbative renormalization group equation. Of course,
this is what had to
be given up to get rid of the Landau pole.
However, the above noted agreement
with the two-loop result for $2.5  \leq R/a \leq 4$
means that numerically
it does agree
with perturbative scaling here.

To compare with the the linear+Coulomb fit as well as to
check the form of the
potential resulting from the new phenomenological running coupling,
a numerical integration
of the new force was performed. A one parameter fit was then
made to the
potential data to determine the constant of integration. This is
shown in Fig.~2 along with the potential data and the
Coulomb+linear fit
of Ref.~[1].
Neither fit used the point at $R/a =2$. The fits are
seen to be nearly
identical from $R/a =4$ to $R/a \simeq 40$. Thus the modified
running coupling
is capable of producing a phenomenologically confining potential
which is
only distinguishable from a linear string-tension
potential at very large
distances.

The two interpretations also give a suprisingly close physical
scale for the lattice spacing. Interpreting the linear term as a
string tension
and using a physical value of $\sqrt{\sigma}=0.44$~GeV
gives $a^{-1}= 6.56$GeV \cite{UKQCD}.
This can be used to determine $\Lambda _L = 9.80$~MeV and
$\Lambda _R = 20.78 \Lambda _ L = 204$~MeV.
Using again the two-loop form from which $\Lambda _R$ is defined
($\Lambda_R = 1/R_0$ for the two-loop force) \cite{Billiore},
and fitting this form to the new force (with parameters fixed
to values previously given) in the short distance
region, $2.5 \leq R/a \leq 4.0$, gives  $R_0/a=30.57$
in the two-loop force.
Thus $a^{-1}=30.57\Lambda _R= 6.22$~GeV, using the
same physical value of
$\Lambda_R$ as above. Thus the two interpretations nearly
agree on the
physical
scale of the lattice.

The same approach was also successfully tried for the SU(3) potential
at $\beta=6/g^2= 6.2$ and $6.4$ using the data of Ref.~[7].
The SU(3) data is not as good statistically as the SU(2) data,
however, nor does
it go to as large a distance, so it is not as stringent a
test as the fits
given above.

What is one to make of the fact that the interquark potential
can be fit
solely to a Coulomb force with a logarithmically running coupling?
First, a true string tension term has
not been ruled out by any means. One can have a string
tension term in
addition
to the modified Coulomb potential, with the Coulomb
potential running
more slowly. Hovever, a fit cannot easily tell the difference
between these.
At the very least what is being shown here is that
unless it can be proven
that the coupling stops running at some accessible distance,
then the current quoted values for the string
tension should be taken
as upper limits, with one-sided error bars of the order of 100\%.
It is therefore
very important to determine the true behavior of the
running coupling in
the intermediate and long distance region. It should also
be pointed out that
the modification introduced by the log-squared term {\em reduces}
the force from the one-loop perturbative value. The problem with the
perturbative force is that it is too strong, not too weak,
at long distances.

Going beyond this, it is interesting to entertain the
possibility of doing
away with the string tension and absolute confinement
altogether for the
continuum Yang-Mills theory and possibly even for QCD. As
mentioned before,
it is not necessary for the linear potential to
extend beyond a few fm to
successfully model meson spectroscopy. The modified
Coulomb potential
could presumably come from summed higher-order perturbation
theory without the need
to invoke non-perturbative physics. Because of the
probable importance
of multiple gluon exchange at long distances, due to the
high effective coupling,
this potential could easily have a substantial
Lorentz-scalar piece, which
may be necessary to obtain the correct heavy-quark spin-orbit
splittings~\cite{scalar}.
Of course {\em lattice} Yang-Mills theories using the Wilson action
necessarily confine at strong coupling, and
thus must have a real string tension
within the region of validity of the strong coupling expansion.
However, it is possible that this region is separated from
the weak coupling region which includes the continuum limit by a
phase transition, as occurs in the U(1) lattice gauge theory.
The scaling
of string tension and pseudo-specific heat are consistent with the
existence of a higher order
phase transition around $\beta=2.5$ for SU(2)
and $\beta=6.7$ for SU(3)~\cite{meZPC}.
This transition would correspond to the remnant of
the finite-temperature
transition which exists on the symmetric lattice~\cite{Eve}.
In other words, it is possible that the conventionally-interpreted
finite-temperature transition is in fact a true four-dimensional
deconfining transition which remains at finite $\beta_c$
as the lattice
size becomes infinite. It should also be mentioned in this vein
that non-compact simulations of SU(2) lattice gauge
theory have failed
to see definite signs of confinement~\cite{NC}. Absolute
confinement
is also questionable for actions
which prohibit negative plaquettes~\cite{MP,me2}.

Finally, the possible effects of dynamical quarks
should be considered.
Even without a fundamental string tension in the pure glue theory,
it is very likely that the color force is strong enough to
cause chiral
symmetry breaking. If the chiral condensate is then polarized
by the strong color
fields surrounding a quark-antiquark pair, a region of
higher than normal
vacuum energy surrounding the pair could be formed~\cite{meZPC}.
This region
can form a kind of bag around the meson
which contributes a linear term to the interquark force,
and may also
contribute to the dynamical mass~\cite{meNCIM}.
A diminishing of the vacuum condensate $<\| \bar{\psi}
\psi \| >$ in the neighborhood of a quark source
has been observed in a lattice
simulation \cite{Feilmair}, lending support to this hypothesis.

\newpage
Fig.~1 SU(2) force data from Ref.~[1] with fits to
one-loop (short-dash),
two-loop (long-dash) and modified-Coulomb (solid line) force.
The Force
and $R$ are in lattice units. Error bars range from about 1/2
the size of
plotted points up to the size of the points.
\vspace*{2ex}

Fig.~2 SU(2) potential data from Ref.~[1]. Dashed line is
Coulomb+linear fit;
solid line is fit to modified Coulomb potential. Error
bars range from 1/10
to 1/3 the size of plotted points.

\end{document}